\documentclass[12pt]{article}
\usepackage{amssymb}
\usepackage{graphicx}
\usepackage[cp1251]{inputenc}
\usepackage{rotating}
\begin{document}
 \bigskip
 \centerline{\bf The Radcliffe Wave is not alone in the Local System}
\bigskip
 \bigskip
 \centerline{\bf
            V. V. Bobylev\footnote [1]{E-mail: bob-v-vzz@rambler.ru},
            A. T. Bajkova
            }
 \bigskip
   \centerline {\small \it Main (Pulkovo) Astronomical Observatory of the Russian Academy of Sciences, St. Petersburg, Russia}
 \bigskip
 \bigskip
The spatial distribution of open star clusters (OSCs) younger than 30 million years old in the Local System was studied. It was shown for the first time that a significant number of OSCs belong to the recently discovered Vela Ridge gas and dust supercloud. The most intriguing property of this sample of OSCs is the presence of periodic perturbations in their vertical coordinates with a maximum amplitude of 47 pc and a wavelength of 1.1 kpc. Thus, the discovered chain of young OSCs is analogous to the Radcliffe Wave, but with a lower amplitude of vertical perturbations, a shorter wavelength, and is, on average, 2 million years older.


\newpage
\section{INTRODUCTION}
The Local System is located between two sections of the spiral arms of the Galaxy's global spiral structure—Perseus and Carina-Sagittarius. The youngest objects belonging to the Local System, such as interstellar dust, molecular clouds, OB stars, T Tauri stars, and the youngest open star clusters (OSCs), exhibit a rich fine structure. The recent availability of high-precision astrometric, spectral, and photometric data facilitates the identification of this structure.

In particular, Alves et al. (2020) first identified the Radcliffe Wave from an analysis of molecular clouds. It is a thin chain of clouds approximately 2.7 kpc long, located between the Sun and the Perseus arm at an inclination to the $Y$ axis of approximately 30$^ \circ$ . The main feature of the Radcliffe Wave is the wave-like behavior of the vertical coordinates with a maximum deviation from the plane of symmetry of the Galaxy of approximately 150 pc. By smoothing a three-dimensional map of the interstellar dust distribution, Kormann et al. (2026) identified a grouping of dust superclouds in the Local System. It included the following seven superclouds: Anguis, Malpolon, Natrix, Radcliffe Wave, Split, Vela Ridge, and Sagittarius Spur Extension. These superclouds lie in the galactic $XY$ plane, inclined to the $Y$ axis at angles from 25$^\circ$ to 35$^\circ$, i.e., they are located almost parallel to each other. It is important to note that in four superclouds, namely, in Malpolon, Natrix, Radcliffe Wave and Vela Ridge, Kormann et al. (2026) found periodic oscillations of the vertical coordinates with wavelengths from 1 to 3~kpc and amplitudes from 30 to 90~pc. In the work of Sorokina et al. (2026), periodic perturbations of the vertical coordinates in the Radcliffe Wave, Malpolon+Natrix and Vela Ridge superclouds were confirmed as a result of processing the three-dimensional dust map of Gontharov et al. (2025).

Studying the spatial and kinematic properties of these superclouds is of great interest in understanding the underlying cause of vertical disturbances. The Radcliffe wave is the most studied in all respects. However, so far there are only a series of hypotheses regarding its origin, none of which is currently generally accepted. Possible scenarios that have been discussed include, for example, the development of a Kelvin-Helmholtz instability in the galactic disk due to the difference in rotational velocities between the dark matter halo and the disk (Fleck 2020); the influence of the Parker instability of the galactic magnetic field as the cause of the formation of wave-like inhomogeneities in the galactic disk (Bobylev et al. 2025a); and the impact of an external impactor on the galactic disk, such as a dwarf galaxy satellite of the Milky Way, a massive dark matter clump, or a globular cluster (Thulasidharan et al. 2022). Options have also been proposed for the impact on the structure of the galactic disk of shock waves from several supernova explosions and their stellar winds, which initiated, for example, the formation of the Local Bubble or the North Polar Spur~ (Marshal, Martin 2023; Konietzka et al. 2024).

The aim of this paper is to study the structure of the Local System using positional data on young open-clusters (OSCs). A fairly reliable connection between young OSCs and the Radcliffe Wave structure has been established (e.g., Konietzka et al. 2024; Bobylev et al. 2025b, 2026). In this paper, we aim to determine whether OSCs are connected with other superclouds, such as Malpolon, Natrix, or Vela Ridge, which exhibit periodic deviations from the Galactic plane of symmetry. For this analysis, we use data from the Hunt and Reffert (2024) catalog.

\section{DATA}
The papers Hunt, Reffert~(2021, 2023, 2024) are devoted to the creation of a catalog of open clusters based on data from the Gaia~(2016) project. In the paper Hunt, Reffert~(2021) a method for identifying cluster members using the HDBSCAN cluster analysis algorithm is described in detail, where the stellar data were taken from the Gaia\,DR2 version (Gaia Collab. 2018). In the papers Hunt, Reffert~(2023, 2024) kinematic and photometric data on stars were taken from the Gaia\,DR3 catalog (Gaia Collab. 2023). The current version of the Hunt, Reffert~(2024) catalog includes 5647 open clusters and 1309 moving groups. Moreover, 3530 open-clusters and 539 moving groups are of high quality, most of them located at heliocentric distances less than 4 kpc. The catalog provides average values of the the OSCs trigonometric parallaxes, proper motions, and radial velocities. Age estimates of the clusters, obtained by these authors using the isochrone method, are also provided.

It is important to note that Hunt and Reffert (2023, 2024) publish distances to clusters calculated using the Gaia parallax zero-point correction of $\Delta\pi=-0.017$~milliarcseconds. This correction value was adopted according to the definition of Lindegren et al. (2021). In this paper, we use just such distances, copied from columns 600–614 of the Hunt and Reffert (2024) catalog and the corresponding heliocentric rectangular coordinates $x,y,z$ from columns 641–690. We select objects with the `o' and `m' marks, i.e., OSCs and moving groups.

\begin{figure}[t]
{ \begin{center}
  \includegraphics[width=0.5\textwidth]{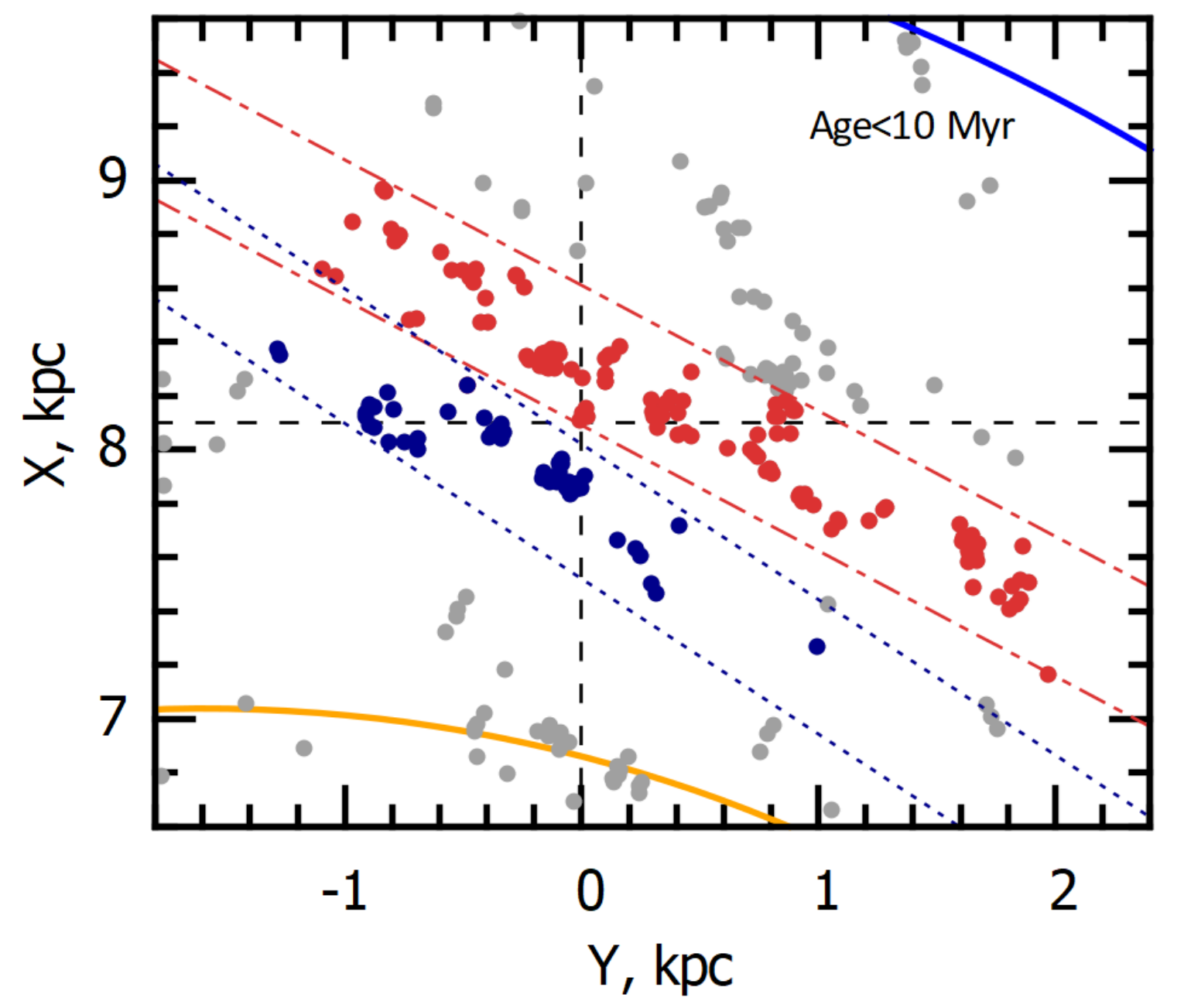}
  \caption{Distribution of OSCs younger than 10 million years projected onto the galactic plane $XY$~--- gray circles, probable members of the Radcliffe Wave structure~--- red circles, probable members of the Vela Ridge structure~--- blue circles, the Sun is located at the point with coordinates $(X,Y)=(8.1,0)$~kpc, selection zones are indicated for each subsample.
  }
 \label{f-1}
\end{center}}
\end{figure}
\begin{figure}[t]
{ \begin{center}
  \includegraphics[width=0.5\textwidth]{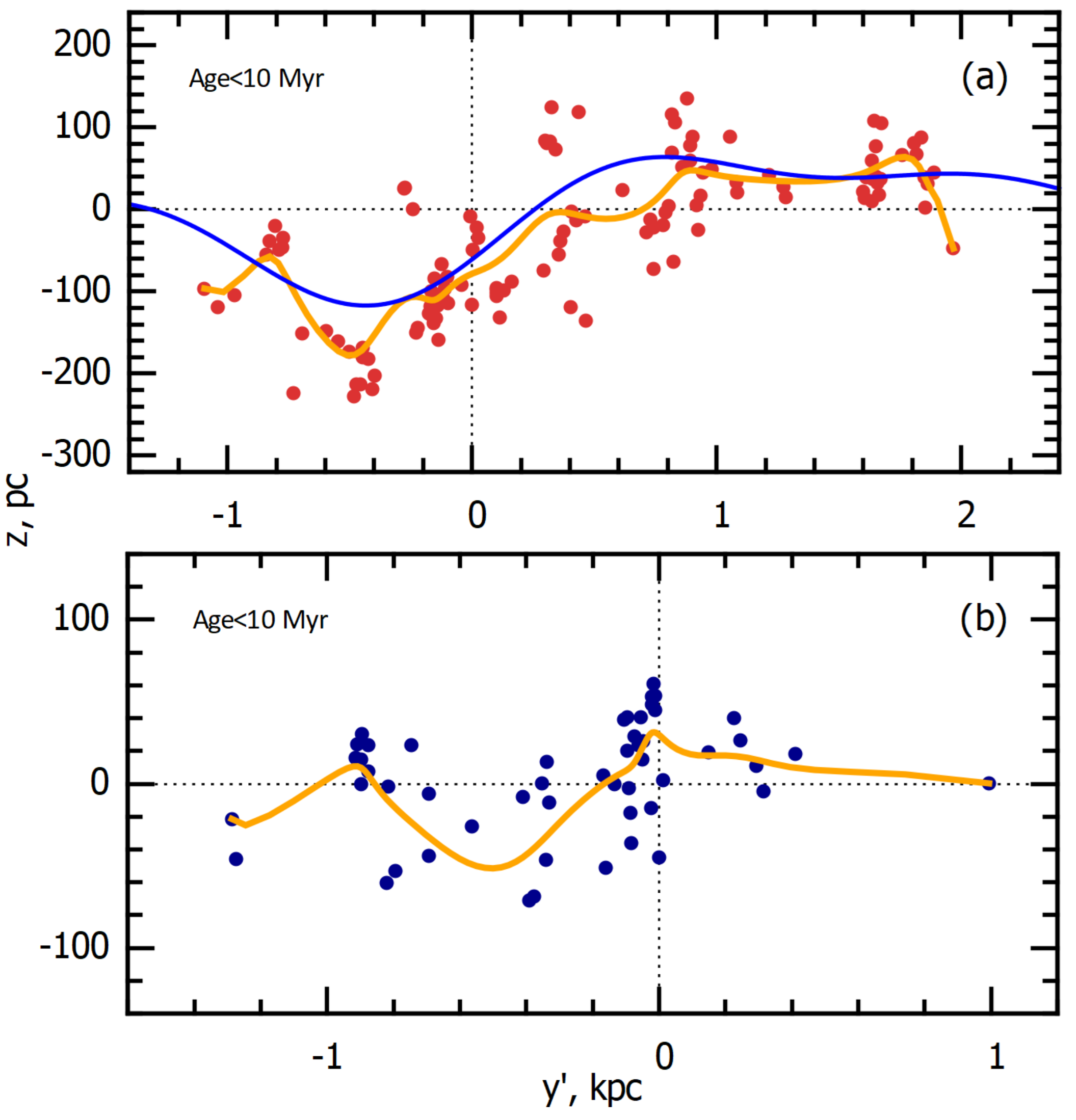}
  \caption{$z$ coordinates versus $y'$ coordinates for probable members of the Radcliffe Wave structure with an age of less than 10 Myr~--- red circles, moving average~--- orange line, found from a similar sample in the work of Bobylev et al. (2025a) based on Fourier analysis of the periodic curve~--- blue line (a); for probable members of the Vela Ridge structure with an age of less than 10 Myr~--- blue circles, moving average~--- orange line (b).
  }
 \label{f-2}
\end{center}}
\end{figure}
\begin{figure}[t]
{ \begin{center}
  \includegraphics[width=0.5\textwidth]{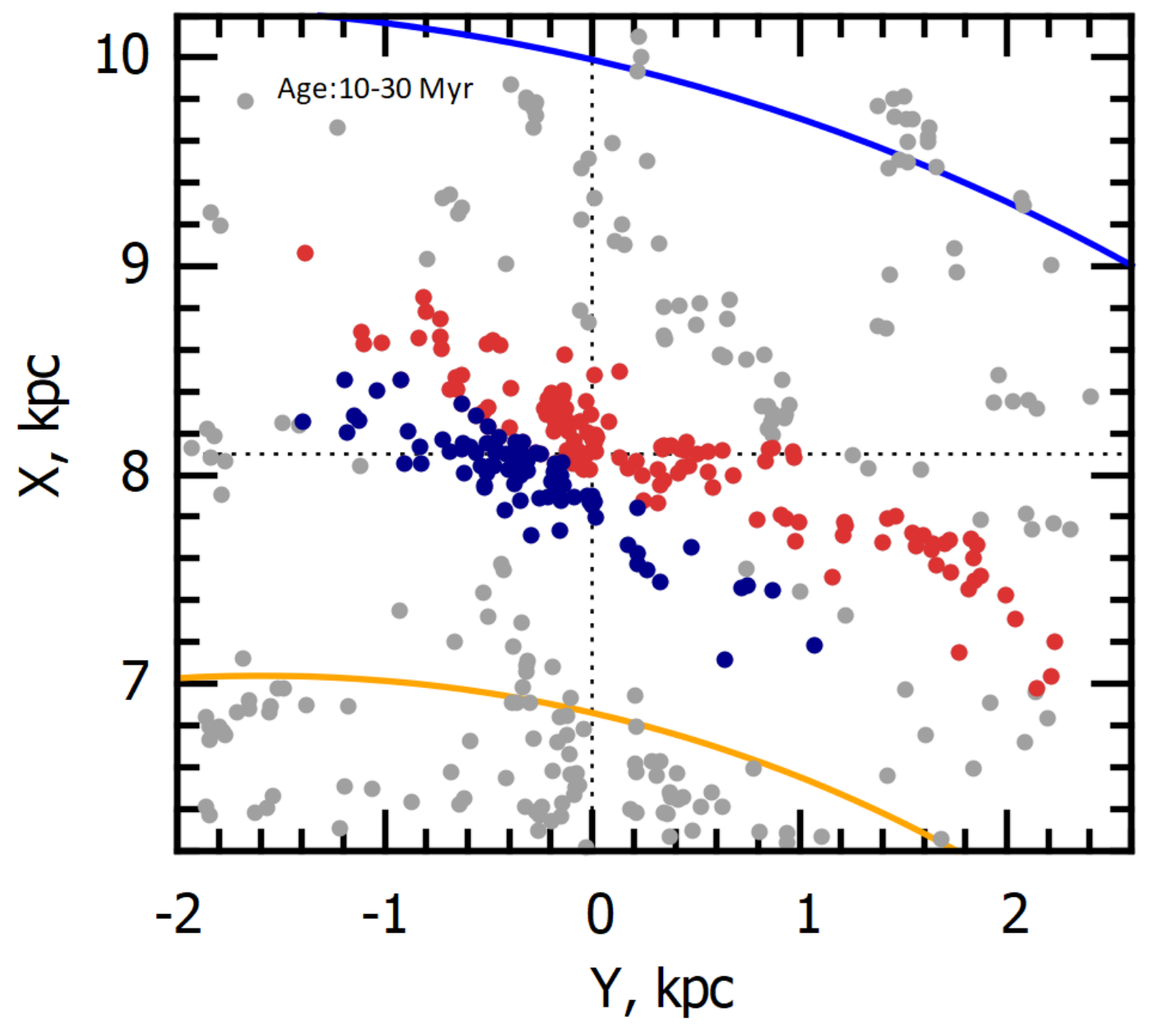}
  \caption{Distribution of OSCs with ages from 10 to 30 million years projected onto the galactic plane $XY$~--- gray circles, probable members of the Radcliffe Wave structure~--- red circles, probable members of the Vela Ridge structure~--- blue circles, the Sun is located at the point with coordinates $(X,Y)=(8.1,0)$~kpc.
  }
 \label{f-3}
\end{center}}
\end{figure}
\begin{figure}[t]
{ \begin{center}
  \includegraphics[width=0.5\textwidth]{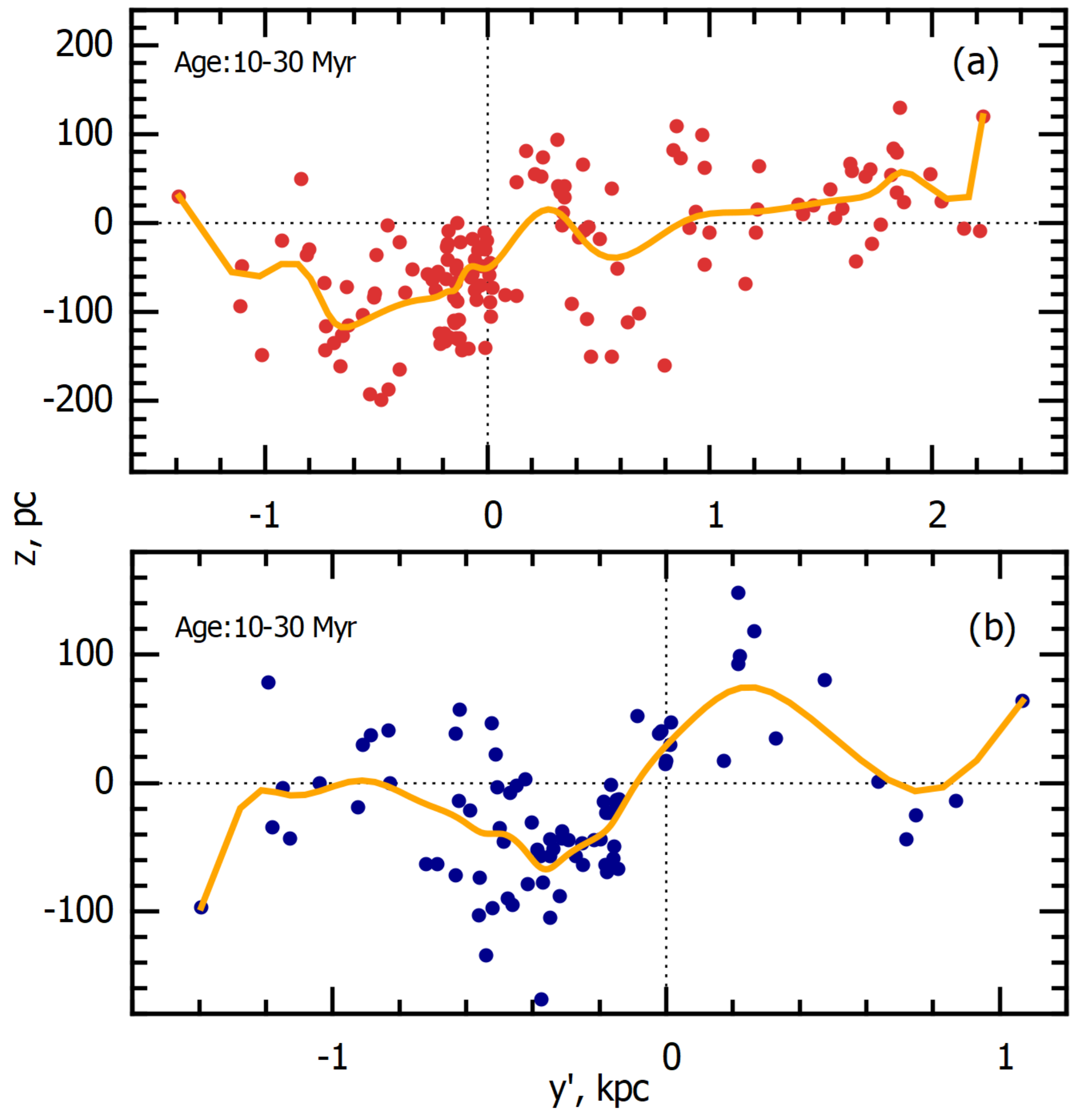}
  \caption{$z$ coordinates versus $y'$ coordinates for probable members of the Radcliffe Wave structure with ages in the range 10-30 Ma~--- red circles, moving average~--- orange line~(a); for probable members of the Vela Ridge structure with ages in the range 10-30 Ma~--- blue circles, moving average~--- orange line~(b).
  }
 \label{f-4}
\end{center}}
\end{figure}
\begin{figure}[t]
{ \begin{center}
  \includegraphics[width=0.5\textwidth]{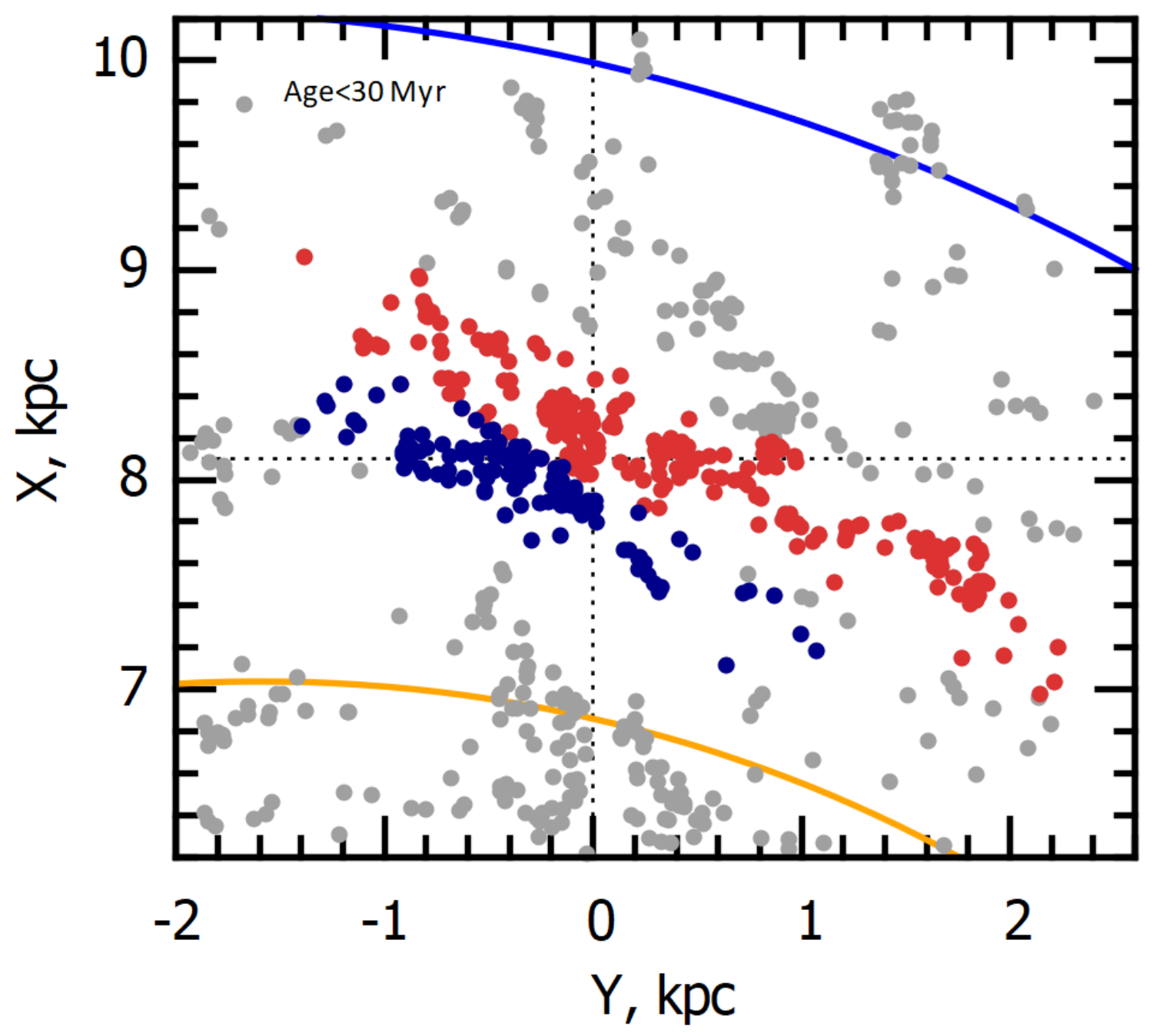}
  \caption{Distribution of OSCs with ages less than 30 million years projected onto the galactic plane $XY$~--- gray circles, probable members of the Radcliffe Wave structure~--- red circles, probable members of the Vela Ridge structure~--- blue circles, the Sun is located at the point with coordinates $(X,Y)=(8.1,0)$~kpc.
  }
 \label{f-5}
\end{center}}
\end{figure}
\begin{figure}[t]
{ \begin{center}
  \includegraphics[width=0.95\textwidth]{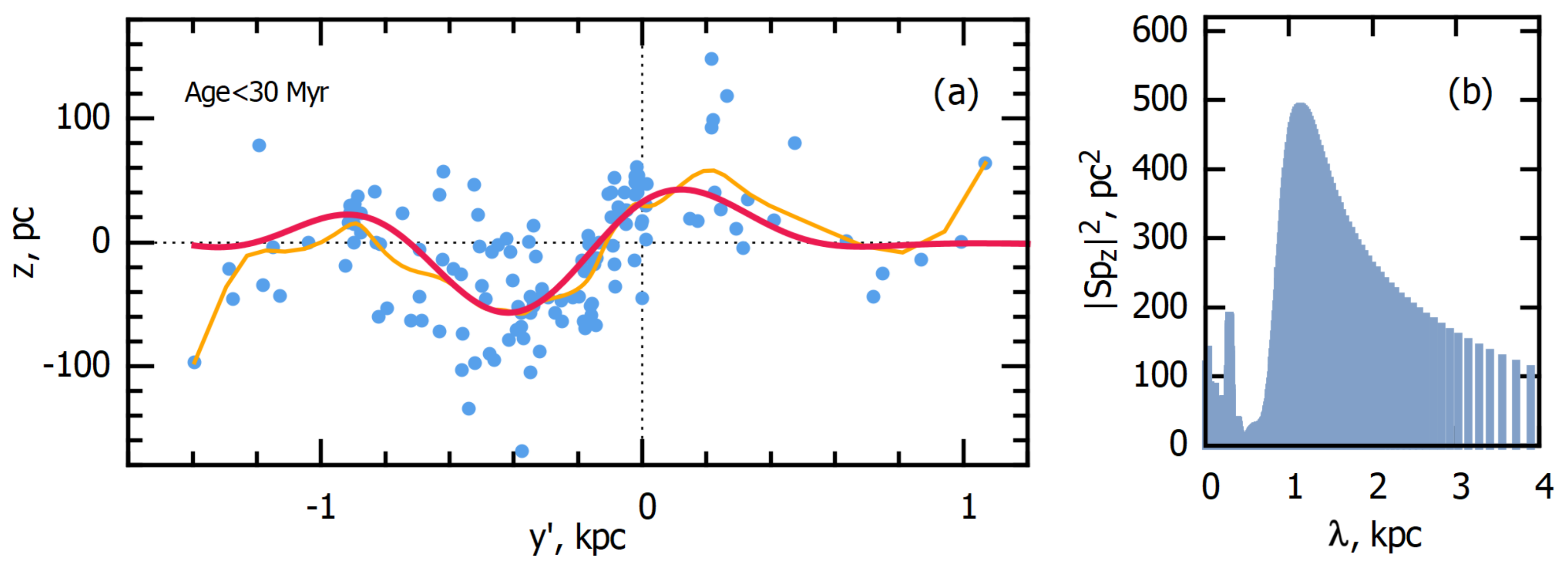}
  \caption{$z$ coordinates versus $y'$ coordinates for probable members of the Vela Ridge structure with ages less than 30 million years~--- blue circles, moving average~--- orange line, periodic curve found based on Fourier analysis~--- red line~(a), power spectrum~(b).
  }
 \label{f-6}
\end{center}}
\end{figure}

  \section*{METHODS}\label{method}
  \subsection*{Coordinate systems}
In this paper, we use various rectangular galactic coordinate systems. One of them is the galactocentric $X,Y,Z$ coordinate system, where the $X$ axis points from the galactic center toward the Sun, the $Y$ axis points in the direction of galactic rotation, and the $Z$ axis points toward the north pole of the galaxy. We adopt the distance from the Sun to the galactic center as $R = 8.1$ kpc, which was derived as a weighted average from a large number of individual estimates in Bobylev and Baykova (2021).

Another coordinate system is the heliocentric $x,y,z$ system, where the $x$-axis points away from the Sun toward the center of the Galaxy, the $y$-axis points in the direction of the Galaxy's rotation, and the $z$-axis points toward the North Pole of the Galaxy. We see that these two coordinate systems differ only in the direction of the $X$ - and $x$-axes.

For the selection and further analysis of the OSCs, we also make the transition to a primed heliocentric coordinate system with the $y'$ axis rotated by an angle $\beta$ as follows
\begin{equation}
 y'= y\cos\beta+x\sin\beta.
 \label{y'-beta}
\end{equation}

 \subsection*{Fourier analysis}
 To study the periodic structure in the coordinates of the OSCs, we use spectral analysis based on the standard Fourier transform of the original sequence $z(y')$:
 \begin{equation}
 \renewcommand{\arraystretch}{1.6}
  \begin{array}{lll}
  \displaystyle
 F(z(y'))= \int z(y') e^{-j{2\pi\over\lambda}y'} dy' = \\\quad
  \displaystyle
  \qquad
 U(\lambda)+jV(\lambda)= A(\lambda) e^{j\varphi(\lambda)},
 \label{F}
 \end{array}
\end{equation}
where
$A(\lambda)=\sqrt{U^2(\lambda)+V^2(\lambda)}$~ is the spectral amplitude, and
$\varphi(\lambda)=\arctan(V(\lambda)/U(\lambda))$~ is the spectral phase.
A distinctive feature of this approach is the search for not just a monochromatic wave with constant amplitude, but a wave that most accurately describes the original data, the spectrum of which coincides with the main peak (lobe) of the calculated spectrum in the wavelength range from $\lambda_{min}$ to $\lambda_{max}$ (within these boundaries, the spectrum smoothly decreases starting from the maximum value, and begins to increase beyond these boundaries).

As a result, we have the desired smooth curve approximating the original data, which is calculated using the inverse Fourier transform formula in the wavelength range we have defined:
\begin{equation}
 \renewcommand{\arraystretch}{2.0}
 z(y')= 2\int^{\lambda_{max}}_{\lambda_{min}} A(\lambda)\cos\biggl( {2\pi y'\over\lambda} + \varphi(\lambda) \biggr) d\lambda.\
 \label{Z}
\end{equation}

  \section{RESULTS}\label{rezult}
 \subsection{OSCs younger than 10 million years}\label{rezult}
 Fig. \ref{f-1} shows the distribution of OSCs younger than 10 million years projected onto the galactic plane $XY$, where probable members of two structures --- Radcliffe Wave and Vela Ridge -- are separately marked. The figure shows two segments of the global four-armed spiral pattern of the Galaxy --- Perseus (blue line in the figure) and Carina-Sagittarius (yellow line in the figure). This pattern with a pitch angle $i=-13^\circ$ is given according to the work of Bobylev and Baykova (2014), and here it is constructed with the value $R_0=8.1$~kpc.

 It should be noted that there are no significant features between the Sun and the Perseus arm that could be associated with the Anguis, Malpolon, and Natrix superclouds, the outlines of which are given in the paper by Kormann et al. (2026). Near the Carina-Sagittarius arm, there is a grouping of OSCs that can be associated with the Sagittarius Spur Extension supercloud (see also Figs. ~\ref{f-3} and \ref{f-5}). However, this supercloud is of no interest to us due to the absence of periodic vertical oscillations in it (Kormann et al. 2026; Sorokina et al. 2026). As a result, in this paper, the main attention is paid to the Radcliffe Wave and Vela Ridge structures.

 For the Radcliffe Wave, the selection zone has a total width of 0.63 kpc, located at an inclination of $\beta=25^\circ$ to the $y$ axis (see the relation~(\ref{y'-beta})). This zone includes 120 OSCs, the average age of the sample is 6.2 Myr. The width of the selection zone of the Vela Ridge structure is narrower, amounting to 0.53 kpc, and has an inclination of $\beta=30^\circ$. This zone includes 54 OSCs with an average age of 6.7 Myr. As can be seen from the figure, in the considered solar neighborhood both selection zones practically do not overlap. Note that the selection zones are not shown in Figs.~\ref{f-3} and \ref{f-5}, although their parameters are identical to the selection zones marked in Fig.~\ref{f-1}.

Fig. \ref{f-2} shows the $z$-coordinates versus the $y'$ -coordinate for probable members of the Radcliffe Wave structure and for probable members of the Vela Ridge structure. This figure shows periodic vertical oscillations for both the Radcliffe Wave (Fig.~\ref{f-2}a) and Vela Ridge (Fig.~\ref{f-2}b). For the Radcliffe Wave, such oscillations are particularly noticeable and have previously been studied using OSCs in the works of Konecki et al. (2024), Bobylev et al. (2025b, 2026). However, we are the first to establish confirmation of periodic vertical oscillations for Vela Ridge using OSCs. However, in constructing Fig. \ref{f-2}b, too few OSCs were used to confidently determine the parameters of the periodic structure. Therefore, similar graphs (Fig.~\ref{f-4}b and Fig.~\ref{f-6}) constructed using a larger number of OSCs in this region are of interest.

The moving averages indicated by the orange lines in Fig. ~\ref{f-2}--\ref{f-6} were plotted automatically using the gnuplot graphics package. The periodic curve shown in Fig.~\ref{f-2}a was obtained by Bobylev et al. (2025a) based on a Fourier analysis using a sample of OSCs with an average age of 5.2 million years, which is virtually identical to that used in the present study to construct the figure.

 \subsection{OSCs with age from 10 to 30 million years}\label{rezult}
 Fig. \ref{f-3} shows the distribution of OSCs with ages from 10 to 30 million years projected onto the galactic $XY$ plane, where probable members of two structures—Radcliffe Wave and Vela Ridge—are separately marked. The Radcliffe Wave selection zone included 144 OSCs, with an average age of 18.9 million years. The Vela Ridge selection zone included 82 OSCs with an average age of 20.8 million years.

Fig. \ref{f-4} shows the $z$ coordinates as a function of the $y'$ coordinate for OSCs with ages from 10 to 30 million years, probable members of the Radcliffe Wave structure, and similarly for probable members of the Vela Ridge structure.

 \subsection{OSCs younger than 30 million years}\label{rezult}
Fig. \ref{f-5} shows the distribution of OSCs younger than 30 million years projected onto the galactic $XY$ plane, with probable members of two structures—Radcliffe Wave and Vela Ridge—marked separately. The Radcliffe Wave selection zone included 264 OSCs, with an average age of 13.1 million years. The Vela Ridge selection zone included 136 OSCs, with an average age of 15.2 million years.

Based on the Fourier analysis of the $z$-coordinates (from $y'$) of 136 OSCs younger than 30 million years, probable members of the Vela Ridge structure, the following estimates of the maximum value of the $z$-coordinate ($z_{max}$ and the wavelength $\lambda$ were obtained:
  \begin{equation}
 \label{sol-Z}
 \begin{array}{lll}
  z_{max}=47.0\pm0.2~\hbox{pc},\\
  \lambda=1.13\pm0.01~\hbox{kpc}.
 \end{array}
 \end{equation}
Error estimates for the sought parameters were found through statistical Monte Carlo simulation. For this, 1000 simulations were conducted using measurement errors $z$ and $y'$.

Figure \ref{f-6}a shows the $z$-coordinates versus the $y'$ -coordinate for 136 OSCs younger than 30 million years, probable members of the Vela Ridge structure, a moving average curve, and a periodic curve found from Fourier analysis. Figure \ref{f-6}b shows the power spectrum. As can be seen from Figure \ref{f-6}a, there is good agreement between the two periodic curves. However, the advantage of Fourier analysis is that this method provides quantitative estimates of the most important characteristics of the wave (the (\ref{sol-Z}) result).

\section{DISCUSSION}\label{rezult}
For the Vela Ridge supercloud, Kormann et al. (2026) found the following vertical disturbance wave parameters: $z_{max}=56.9$~pc and $\lambda=3.079$~kpc. Unlike other superclouds, Vela Ridge is the most loose and consists of separate pieces, as can be seen, for example, in Fig. 4 of these authors' paper. The largest piece is located in the region of  $(X,Y)\approx(8,-1)$ ~kpc.

When analyzing the Vela Ridge supercloud based on the three-dimensional map of the distribution of absorbing matter by Gontcharov et al. (2025), the following parameters of the vertical disturbance wave were found in the work of Sorokina et al. (2026): $z_{max}=51$~pc and a shorter wavelength $\lambda=1.84$~kpc.

As can be seen from (\ref{sol-Z}), different approaches~--- for dust and for OSCs yielded close estimates of the amplitude of vertical disturbances $z_{max}$. The values of the wavelength $\lambda$ differ significantly.

It is interesting to note that when selecting OSCs in the Vela Ridge supercloud region, their average age is approximately 2 million years older than in the Radcliffe Wave region.

In addition to vertical oscillations, the Radcliffe Wave as a whole moves radially toward the galactic anticenter and also shifts in the direction of galactic rotation. This was first established by Konecki et al. (2024) based on an analysis of open clusters and confirmed by Bobylev et al. (2025b), also based on open clusters. Furthermore, Bobylev et al. (2025b) showed that in the Radcliffe Wave selection zone, younger open clusters are located, on average, slightly further from the galactic center compared to older clusters.

A comparison of all these features suggests that the chain of OSCs belonging to the Vela Ridge supercloud is analogous to the Radcliffe Wave. The OSCs belonging to the Vela Ridge supercloud are located slightly closer to the galactic center and are, on average, 2 million years older than the Radcliffe Wave. These two structures are spatially close to each other but do not intersect. They represent a single space-wave process, extended in time. Apparently, both structures were formed by a single mechanism. The nature of this mechanism is still unknown.

  \section{CONCLUSION}\label{concl}
The spatial distribution of open star clusters younger than 30 million years old in the Local System was studied. Data for the analysis, such as estimated OSCs coordinates and ages, were taken from the Hunt and Reffert (2024) catalog.

For the first time, young clusters belonging to the recently discovered Vela Ridge supercloud have been demonstrated. The most intriguing property of this sample of OSCs is the presence of periodic perturbations in their vertical coordinates.

Based on a sample of 136 such OSCs with ages less than 30 million years, the following parameters of this periodicity were found using Fourier analysis: the maximum amplitude is $47.0\pm0.2$~pc, and the wavelength is $1.13\pm0.01$~kpc. Just as in the case of the Radcliffe wave, this periodicity is not a monochromatic wave (like a simple sine wave), but is a damped process.

\medskip

 \bigskip{BIBLIOGRAPHY}\medskip{\small
\begin{enumerate}

 \item
J. Alves, C. Zucker, A. A. Goodman, et al., Nature {\bf 578}, 237 (2020).

 \item
V.V. Bobylev and  A.T. Bajkova, MNRAS {\bf 437}, 1549 (2014). 

 \item
V.V. Bobylev, A.T. Bajkova, Astron. Rep. {\bf 65}, 498 (2021). 

 \item
V.V. Bobylev, N.R. Ikhsanov, A.T. Bajkova, Astron. Rep. {\bf 69}, 786 (2025a).

 \item
V.V. Bobylev, N.R. Ikhsanov, A.T. Bajkova, Astrophys. Bull. {\bf 80}, 181 (2025b).

 \item
V.V. Bobylev,  A.T. Bajkova, N.R. Ikhsanov, Astrophys. Bull. {\bf 81}, Issue~3 (2026).

\item
R. Fleck), Nature {\bf 583}, 24 (2020).

\item
Gaia Collab, T. Prusti, et al., Astron. Astrophys. {\bf 595}, 1 (2016). 

 \item
Gaia Collab, A.G.A. Brown,  et al., Astron. Astrophys. {\bf 616}, 1 (2018). 

 \item
Gaia Collab, A. Vallenari, et al.,  Astron. Astrophys. {\bf 674}, 1 (2023). 

 \item
G. A.  Gontcharov, A. A. Marchuk, S. S. Savchenko,  et al.,
 Res. Astron. Astrophys. {\bf 25}, No 12, id. 125016  (2025).

\item
E.L. Hunt, S. Reffert, Astron. Astrophys {\bf 646}, A104 (2021).

\item
E.L. Hunt, S. Reffert, Astron. Astrophys {\bf 673}, A114 (2023).

\item
E.L. Hunt, S. Reffert, Astron. Astrophys {\bf 696}, A42 (2024).

\item
R. Konietzka, A.A. Goodman, C. Zucker, et al.,  Nature {\bf 628}, 62 (2024).

\item
L. A. Kormann, J. Alves, M. P. Gonz\'alez, et al., Astron. Astrophys. {\bf 706}, A161 (2026).

\item
A. Marchal, P.G. Martin, Astrophys. J. {\bf 942}, 70 (2023).

\item
V. I. Sorokina, V. V. Bobylev, G. A. Gontcharov, and A. T. Bajkovav, arXiv:2607.14551 (2026).

\item
L. Thulasidharan, E. D'Onghia, E. Poggio, et al., Astron. Astrophys. {\bf 660}, 12 (2022).

\end{enumerate} }
\end{document}